\documentclass[a4paper,11pt]{article}
\usepackage{jcappub} % for details on the use of the package, please see the JINST-author-manual
\DeclareMathOperator{\sech}{sech}
\usepackage{subfigure}
\title{Constraining symmetron fields with a levitated optomechanical system}
\author{Jiawei Li}
\author[1]{and Ka-Di Zhu\note{Corresponding author.}}
\affiliation{Key Laboratory of Artificial Structures and Quantum Control (Ministry of Education), School of Physics and Astronomy, Shanghai Jiao Tong University,\\
DongChuan Road,
Shanghai 200240, China}

\emailAdd{zhukadi@sjtu.edu.cn}

\abstract{The symmetron, one of the light scalar fields introduced by dark energy theories, is thought to modify the gravitational force when it couples to matter. However, detecting the symmetron field is challenging due to its screening behavior in the high-density environment of traditional measurements. In this paper, we propose a scheme to set constraints on the parameters of the symmetron with a levitated optomechanical system, in which a nanosphere serves as a testing mass coupled to an optical cavity. By measuring the frequency shift of the probe transmission spectrum, we can establish constraints for our scheme by calculating the symmetron-induced influence. These refined constraints improve by 1 to 3 orders of magnitude compared to current force-based detection methods, which offer new opportunities for the dark energy detection.}

\begin{document}
\maketitle
\flushbottom

\section{Introduction}\label{1}

Observations on the cosmic scale have indicated an accelerating expansion of our universe, but a convincing explanation of this phenomenon remains elusive\cite{Brax_2018,Riess_1998,Schmidt_1998}. A critical issue is that there is a discrepancy between different measurements of Hubble constant $H_0$ \cite{PhysRevD.100.043537,LOMBRISER2019134804,Velten_2014,Solà_2013}. The Planck satellite gives the value $H_0 = (67.8 \pm 0.5)$km/s$\cdot$Mpc\cite{refId0} while the SH0ES team result indicates $H_0 = (73.0 \pm 1.0)$km/s$\cdot$Mpc\cite{Riess_2022}. Common explanations include simply systematic errors, underdensity-area-induced expansion acceleration\cite{refId02,10.1093/mnras/stac396} and new physics beyond the standard model\cite{SCHONEBERG20221,RevModPhys.90.025008}. However, the first explanation fails because the disagreement is sufficiently large across different statistical approaches \cite{10.1093/mnras/sty418,Cardona_2017} and the second one is ruled out by observations\cite{10.1093/mnras/stab3077}. The most popular answer seems to be dark energy, where the acceleration of expansion is explained within the framework of the scalar field\cite{doi:10.1142/S021827180600942X}. Additionally, high energy physics theories suggest that the coupling between the scalar field and matter results in the so-called fifth force beyond the four fundamental forces, which leads to the modification of the gravitational force \cite{PhysRevD.108.124050}.

Considering the fact that the scalar fields evade the search in local environment, there must be a screening mechanism to shield them from laboratory detection\cite{JOYCE20151}. Several types of screening mechanism have been proposed, including the chameleon\cite{Burrage_2015,Qvarfort_2022}, the symmetron\cite{PhysRevD.86.044015,PhysRevD.84.103521}, K-mouflage\cite{PhysRevD.90.023507} and Vainshtein\cite{VAINSHTEIN1972393}. Among what is discussed above, the former two share a unified description but different Lagrangians. They both mediate long-range forces in low density areas. But in high density areas, the mass of the chameleon increases and the mediated force become short-range\cite{Burrage_2015}, while the symmetron develops nonlinearity that results in its decoupling from matter\cite{PhysRevD.96.124029}. In contrast, the latter two are mainly determined by the steep field gradients of the scalar field and the second derivative of the field, respectively\cite{doi:10.1142/S0218271809016107,PhysRevD.79.064036,PhysRevD.101.064065}.

Here, we focus on the symmetron model. To be specific, the symmetron model relies on the vacuum expectation value of a scalar field , which undergoes a  $\mathbb{Z}_2$  symmetry breaking transition as the local matter density decreases from a critical value\cite{Burrage_2016,Brax_2013,PhysRevD.72.043535}. In contrast, when the matter density is large, the effective field potential has only one minimum and remains zero. There have been several tests and related analyses of constraints on the symmetron field, including satellite-based tests\cite{f6f4498b01ee424197429ba2efeea1b8}  , torsion pendulum experiments\cite{PhysRevLett.98.021101,PhysRevLett.110.031301}, Casimir-force detection\cite{PhysRevD.75.077101,PhysRevLett.78.5,PhysRevD.101.064065}, neutron tests\cite{article,article1} and atomic interferometry\cite{Burrage_2015,PhysRevD.94.104069}. With the results of the mentioned tests, the physically plausible parameter space has been covered by considering different constraints\cite{Burrage_2016,article2}.

In this paper, we propose a scheme aimed for probing the symmetron-force gradient with an optically trapped  nanosphere for future experiments. The article is organized as follows: In Sec.~\ref{2}, we present the theoretical model, including a brief introduction of symmetron field as well as a solution considering the screening mechanism. We also derive the relation between the symmetron force gradient and the force-induced frequency shift via quantum optics theory. In Sec.~\ref{3}, we determine the minimum measurable frequency shift, considering the spectroscopic performance and dominant noise processes. Additionally, we establish the constraints on symmetron through the aforementioned calculation and demonstrate an improvement of 3 to 1 orders of magnitude for $\mu$ in the range  $10^{-1}$eV to $ 10^{-4}$eV. In Sec.~\ref{4}, we summarize the paper and discuss potential future research directions.

\section{Model and theory}\label{2}
\subsection{The cavity optomechanical system}\label{2.1}
\begin{figure}[ht]
\centering\includegraphics[width=7cm]{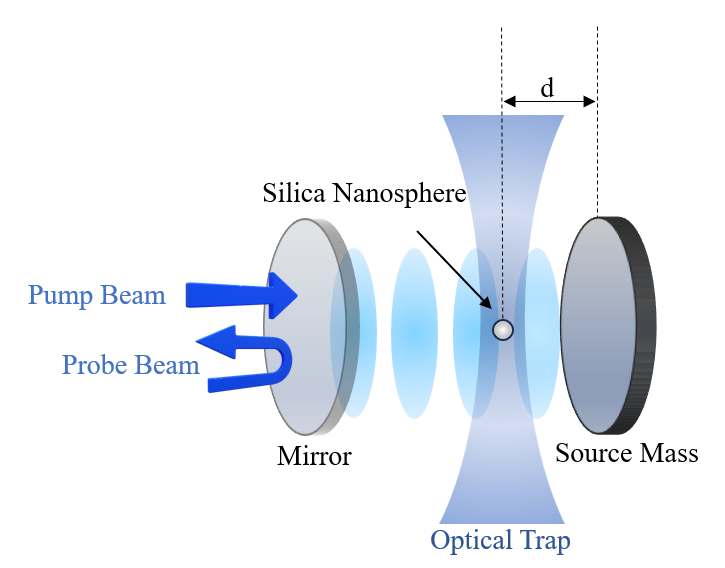}
\caption{\label{fig:2}The schematic diagram of our system which consists of an optically trapped  fused silica nanosphere driven by a pump beam and a weak probe laser, with radius $R=100$nm and density $\rho_{sphere}=2.2\times10^3$kg$/m^3$. The system includes a transflective mirror as well as another mirror served as source mass. The sphere is cooled in the cavity and positioned at a  distance $d$ from the source mass surface. }
\end{figure}

Our scheme is based on a system shown in Fig.~\ref{fig:2}, composed of a fused silica sphere trapped in an optical cavity with resonance frequency $\omega_{cav}$. we utilized a pump light field with frequency $\omega_p $ and a weak probe field with frequency $ \omega_s$\cite{PhysRevD.106.095007,PhysRevLett.105.101101,PhysRevD.95.044014}. Generally, we have the Hamiltonian of the entire system in a rotating frame given by \cite{PhysRevA.77.033804,PhysRevA.63.023812}
\begin{eqnarray}
H  = \hbar\Delta_{p}c^{\dagger}c + \hbar\omega_{n}a^{\dagger}a - \hbar g c^{\dagger}c(a^{\dagger} + a) - i\hbar\Omega_{p}(c - c^{\dagger}) - i\hbar\Omega_{s}(ce^{i\delta t} - c^{\dagger}e^{-i\delta t}),
\label{eq:six}   
\end{eqnarray}
where $\Delta_p = \omega_{cav} - \omega_p$ is pump-cavity detuning which we set to $\Delta_p =0$ for the this paper; $c(c^{\dagger})$ and $a(a^{\dagger})$ denote the annihilation (creation) operator of the cavity and the bosonic annihilation (creation) operator of the quantum harmonic oscillator; the interaction term $H_{\text{int}} = \hbar g c^{\dagger}c(a + a^\dagger)$ describes the coupling between the cavity and the oscillation of the trapped particle with the coupling strength $g$; $\delta=\omega_s-\omega_p$ is the probe-pump detuning; $\Omega_p$ and $\Omega_s$ are the Rabi frequency for the pump field and the probe field in the cavity respectively, which are proportional to $\sqrt{P_p}$ and $\sqrt{P_s}$. Here $P_p$ and $P_s$ are the power of the pump field and the probe field. 
Moreover, the system measures the force gradient with the optomechanical setup as follows: When under a uniform force, the nanosphere experiences a harmonic trapping potential with spring constant $k$ defined by $\omega_n = \sqrt{k/m}$, where $\omega_n$ is the resonance frequency. However, when it experiences a non-vanishing force gradient, the spring constant is modified to $k' = k +\partial F/\partial x$, leading to a change of frequency
\begin{eqnarray}
\omega'_n \approx \sqrt{\frac{k}{m} \left(1 + \frac{\partial F/\partial x}{k} \right)}.
\label{eq:seven}   
\end{eqnarray}

Given $\omega_n \approx \omega'_n$,we have
\begin{eqnarray}
\frac{\partial F/\partial x}{2k} \approx \frac{\omega'_n}{\omega_n} - 1,
\label{eq:eight}   
\end{eqnarray}
and the relation between the frequency shift $\Delta \omega$ and the force gradient $\frac{\partial F}{\partial x}$ is
\begin{eqnarray}
\frac{\partial F}{\partial x} = \frac{2k\Delta \omega}{\omega_n}.
\label{eq:nine}   
\end{eqnarray}

Then we define the operator $N=a+a^{\dagger}$, which is related to the position of the quantum oscillator. Using the Heisenberg equation and the communication relations that $[c,c^\dagger] = 1$ and $[a,a^\dagger] = 1$, we have the evolution of $c$ and $N$ described by the quantum Langevin equations with additional  damping terms:
\begin{eqnarray}
\frac{dc}{dt} =- (i\Delta_p + \kappa) c + igNc + \Omega_p + \Omega_s e^{-i\omega t}+ \sqrt{\kappa_{ex}}\hat{a}_{in} + \sqrt{\kappa_0}\hat{f}_{in},
\label{eq:ten}   
\end{eqnarray}

\begin{eqnarray}
\frac{d^2N}{dt^2} + \gamma_n \frac{dN}{dt} + \omega_n^2 N = 2\omega_n gc^{\dagger}c + \hat{\xi},
\label{eq:eleven}   
\end{eqnarray}
here $\kappa = \kappa_0 + \kappa_{\text{ex}}$ and $\gamma_n$ are the energy damping rate of the cavity and the damping rates of the vibrational mode of the nanosphere, respectively. $\hat{a}_{\text{in}}(t)$ is the Langevin noise operator, which has mean value of $\langle \hat{a}_{\text{in}}(t) \rangle = 0$ with correlation function written as $\langle \hat{a}_{\text{in}}(t) \hat{a}_{\text{in}}(t_0) \rangle \sim \delta(t - t_0)$. And $\hat{f}$ has similar form of correlators as those for $\hat{a}_{\text{in}}$. The operator $\hat{\xi}$ stands for the effects of the thermal bath from the non-Markovian stochastic process and the Brownian. It also has a zero mean value with correlation function is given by\cite{PhysRevA.63.023812,gardiner2004quantum}  

\begin{eqnarray}
\left\langle \hat{\xi(t)^{\dagger}}\hat{\xi(t')} \right\rangle = \frac{\gamma_n}{\omega_{n}} \int\frac{d\omega}{2\pi} 
\omega e^{-i\omega(t-t')} 
\left[ 1 + \coth\left( \frac{\hbar\omega}{2k_BT} \right) \right], 
\label{eq:twelve}   
\end{eqnarray}
where $k_B$ is the Boltzmann’s constant and $T$ is the effective resonator temperature. Each Heisenberg operator can be separated into its steady-state and fluctuation part:
\begin{eqnarray}
c=c_0+\delta c,
\label{eq:13}   
\end{eqnarray}
\begin{eqnarray}
N=N_0+\delta N.
\label{eq:14}   
\end{eqnarray}
Then we can get the steady-state solutions of Eqs.~(\ref{eq:ten}) and (\ref{eq:eleven}) as

\begin{eqnarray}
c_0 = \frac{\Omega_p}{(i\Delta_{pu} + \kappa) - igN},
\label{eq:15}   
\end{eqnarray}

\begin{eqnarray}
N_0 = \frac{2g|c|^2}{\omega_n}.
\label{eq:16}   
\end{eqnarray}

Taking Eqs.~(\ref{eq:15})and (\ref{eq:16}) into the Langevin equations, one can get the linearized Langevin equations by dropping the noise terms and nonlinear terms since we choose a weak driving field.
 
\begin{eqnarray}
\langle \delta \dot{c} \rangle = -\kappa \langle \delta c \rangle + ig(N_0 \langle \delta c \rangle + c_0\langle\delta N\rangle) 
+ \Omega_p + \Omega_s e^{-i\delta t},
\label{eq:17}   
\end{eqnarray}

\begin{eqnarray}
\langle\delta\Ddot{  N}\rangle + \gamma_n \langle\delta\dot{N}\rangle + \omega^2_n \langle \delta N\rangle = 2\omega_n gc^{2}_0.
\label{eq:18}   
\end{eqnarray}

To solve these equations, we introduce the following ansatz\cite{10.5555/1817101}:

\begin{eqnarray}
\langle \delta c \rangle = c_+ e^{-i \delta t} + c_- e^{i \delta t},
\label{eq:19}   
\end{eqnarray}

\begin{eqnarray}
\langle \delta N \rangle = N_+ e^{-i \delta t} + N_- e^{i \delta t}.
\label{eq:20}   
\end{eqnarray}

By substituting these expressions into the Eqs.~(\ref{eq:ten}) and (\ref{eq:eleven}), we obtain the solutions of interest:

\begin{eqnarray}
c_+ = \frac{\Omega_s[B(E - P) + G\omega_0]}{B(E^2 - P^2) + 2PG\omega_0},
\label{eq:21}   
\end{eqnarray}

\begin{eqnarray}
\Omega^2_p = \omega_0(\kappa^2 + (\Delta_p - g^2\omega_0/\omega_n)^2).
\label{eq:22}   
\end{eqnarray}

The parameters in the above equations are given as follows:$E = \kappa - i\delta$, $P = i\Delta p - igN_0$, $B = -\delta^2 - i\gamma_n\delta + \omega_n^2$, $G = 2ig^2\omega_n$, $N_0 = 2g|c_0|^2/\omega_n$, and $\omega_0 = |c_0|^2$ can be resolved by Eq.~(\ref{eq:22}). Here $c_+$ is a parameter analogous to the linear optical susceptibility. Its real part and imaginary part exhibit absorptive and dispersive behavior, which can be observed by detecting the transmission spectrum of the probe field. 

To calculate the transmission of the probe field, we first calculate the mean value of output field by using the input-output relation 
\begin{eqnarray}
c_{\text{out}}(t) = c_{\text{in}}(t) - \sqrt{2\kappa} c(t), 
\label{eq:ad4}
\end{eqnarray}
where $c_{\text{in}}$ and $c_{\text{out}}$ are the input and output operators; thus, we can obtain the expectation value of the output field as\cite{bowen2015quantum}:

\begin{eqnarray}
\langle c_{out}(t) \rangle =( \frac{\Omega_p}{\sqrt{2\kappa}}-\sqrt{2\kappa} c_0)e^{-i\omega_pt}+ (\frac{\Omega_s}{\sqrt{2\kappa}}-\sqrt{2\kappa}c_+) e^{-i(\omega_p+\delta)t}-\sqrt{2\kappa}c_-e^{-i(\omega_p-\delta)t}.
\label{eq:23}   
\end{eqnarray}

%with
%\begin{eqnarray}
%c_-=
%\label{eq:24}   
%\end{eqnarray}

Focus on the component with frequency $\omega_p +\delta= \omega_s$, we deduce the transmission of the probe field, defined as the ratio of the output field and input field at the probe frequency:

\begin{eqnarray}
\mathcal{T}(\omega_s) = \frac{\Omega_s /\sqrt{2\kappa}-\sqrt{2\kappa}c_+}{\Omega_s/ \sqrt{2\kappa}}. 
\label{eq:24}   
\end{eqnarray}

\subsection{The symmetron force}\label{2.2}
First of all, we have the Lagrangian density of the symmetron field in natural unit($c=\hbar=1$) 
\begin{equation}
\mathcal{L}=-\frac{1}{2}(\partial\phi)^{2}-\frac{1}{2}(\frac{\rho}{M^{2}}-\mu^{2})\phi^{2}-\frac{1}{4}\lambda\phi^{4}-\frac{\mu^{4}}{4\lambda}.
\label{eq:one}   
\end{equation}
The parameters in the above equation have the following meanings: $\mu$, serves as the tachyonic mass; $\lambda$ is a dimensionless self-coupling constant; and $M$ is a parameter characterizing the coupling strength to matter with the dimension of mass. Considering the interaction with background matter of density $\rho$, we get a $\mathbb{Z}_2$  symmetry breaking potential as
\begin{equation}
V_{eff}(\phi)=\frac{1}{2}(\frac{\rho}{M^{2}}-\mu^{2})\phi^{2}+\frac{1}{4}\lambda\phi^{4}+\frac{\mu^{4}}{4\lambda}.
\label{eq:two}   
\end{equation}

It's obvious that there is only one minimum of the effective potential at $\phi = 0$ when the background matter density is higher than the critical value $\rho_{\text{crit}} = \mu^2 M^2$. In contrast, as the matter density decreases to 0, the minima of the effective potential split into two values at $\phi_{\text{min}} = \pm{v}$ with $v = \mu/\sqrt{\lambda}$. According to the nature of symmetron field that it has two minima for effective potential, a unique phenomenon occurs which is not present in other chameleon models. if the Compton wavelength is smaller than the size of the vacuum cavity, it would be possible for the symmetron field to form a domain wall during the process pumping out the air from the cavity\cite{Burrage_2016}. It is because that the field appears to emerge at either minimum with the equal probability. Here, we only consider the positive value, and then we can get the equation of motion and the force which a test matter $m_{test}$ in the symmetron field experiences with Eq.~(\ref{eq:two})
\begin{equation}
\Vec{\nabla}^{2}\phi=(\frac{\rho}{M^{2}}-\mu^{2})\phi+\lambda\phi^{3},
\label{eq:three}   
\end{equation}
\begin{equation}
F=-\frac{m_{test}}{M^{2}}\phi\Vec{\nabla}\phi.
\label{eq:four}   
\end{equation}

To solve the equation of motion , we consider a monotonically rising field towards the vacuum expectation value as follows(see refs.~\cite{PhysRevD.97.064015,PhysRevD.99.024045,PhysRevD.80.104002} for detailed treatment):
\begin{equation}
\phi(x)=v\tanh(\frac{\mu x}{\sqrt{2}}),
\label{eq:1.1}   
\end{equation}
where $x$ is the distance from the surface of the plate to the center of the sphere.

In our scheme, we utilize a sphere-plate system, in which we should consider the so-called screening factor $\lambda_{sphere}$ for the nanosphere we use here.The screening factor characterizes the behavior of the symmetron field when the matter is sufficiently large or dense that $\rho \geq \mu^2 M^2$. In that case, the interaction is proportional to only a small fraction  $\lambda_{sphere} \ll 1$ of the whole matter. So the Eq.~(\ref{eq:four}) is modified by replacing $m_{test}$ with $\lambda_{sphere} m_{test}$ \cite{PhysRevLett.104.231301}
\begin{equation}
F=-\frac{\lambda_{sphere} m_{test}}{M^{2}}\phi\Vec{\nabla}\phi.
\label{eq:1.2}   
\end{equation}

where we introduce the screening factor for a sphere with radius $R$ as\cite{PhysRevD.101.064065}
\begin{equation}
\lambda_{sphere}\approx min(\frac{3M^{2}}{\rho R^{2}},1).
\label{eq:1.3}   
\end{equation}
The above form of $\lambda$ allows the force to have a non-vanishing prefactor at the limit when $\lambda_{sphere} \rightarrow 0$ as
\begin{equation}
\frac{\lambda_{sphere} m_{sphere}}{M^{2}}=4\pi R.
\label{eq:1.4}   
\end{equation}

Take Eqs.~(\ref{eq:1.1}) and (\ref{eq:1.4}) into Eq.(~\ref{eq:1.2}), the symmetron force and its gradient can be written as 
\begin{equation}
F=-\frac{4\pi v^{2}\mu R}{\sqrt{2}}\tanh(\frac{\mu x}{\sqrt{2}})\sech^{2}(\frac{\mu x}{\sqrt{2}}),
\label{eq:1.5}   
\end{equation}

\begin{equation}
\frac{\partial F}{\partial x}=\frac{2\pi \mu^4 R}{\lambda} \left(3\tanh^{2}\frac{\mu x}{\sqrt{2}}-1\right) \sech^{2}\left(\frac{\mu x}{\sqrt{2}}\right).
\label{eq:1.6}   
\end{equation}

It should be emphasized that the density of the sphere and mirrors is  sufficiently large, so the symmetron field is  zero everywhere inside these objects. Therefore, we focus only on the nonvanishing field in the vacuum.

\begin{figure}[ht!]
	\centering
	\subfigure{
		\includegraphics[width=7cm]{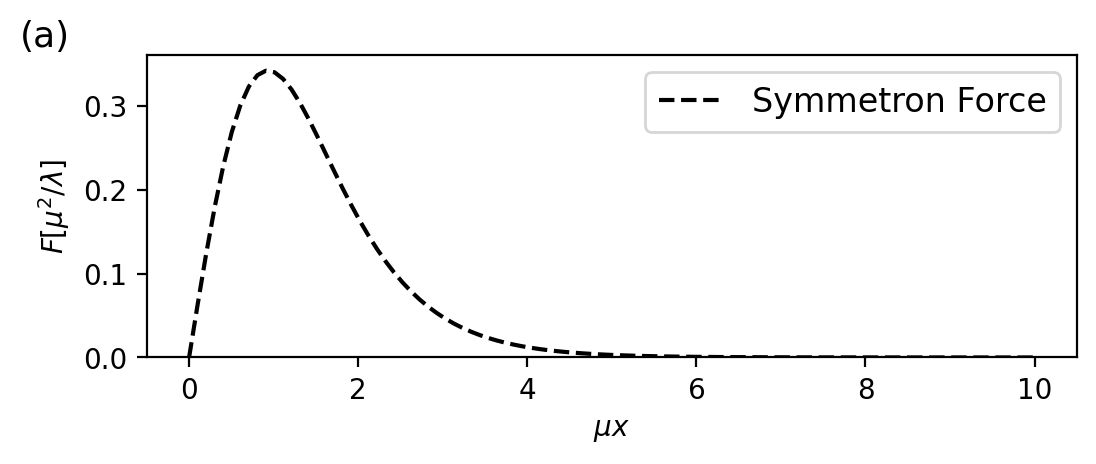}
        \label{fig:1-1}
        \vspace{0.02mm}
}
    \subfigure{
		\centering
		\includegraphics[width=7cm]{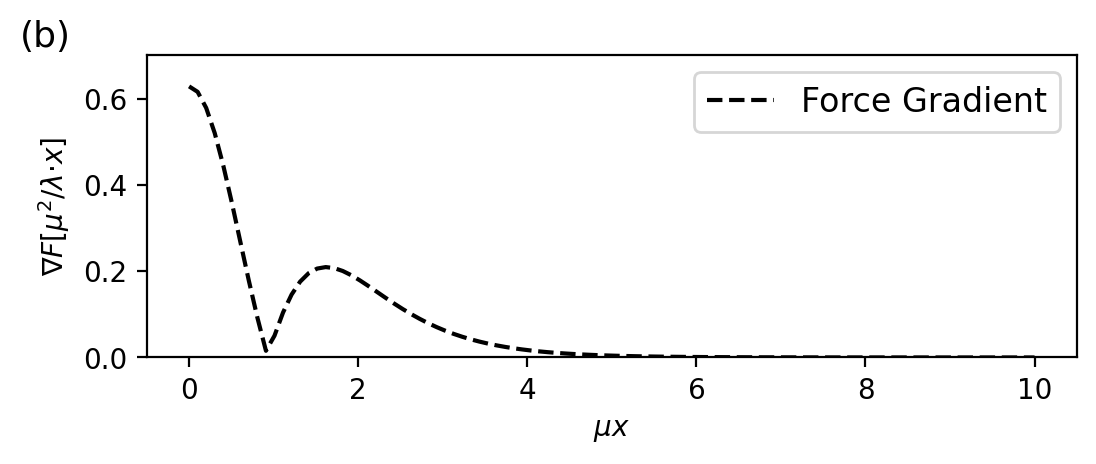}
        \label{fig:1-2}
}
\caption{\label{fig:1}The symmetron force (a) and the force gradient(b) plotted as functions of the distance from the plate to the sphere. To be convenient, we choose $\mu^2/\lambda$ and $\mu^2/\lambda \cdot x$ as dimension, respectively. The criteria of small ball is $\mu R \ll 1$, by which we can ignore influence of sphere on the symmetron field from the plate. Here, we choose $\mu R$=0.1 to give an example of the symmetron mechanism.}
\end{figure}

In Fig.~\ref{fig:1}, we plot the symmetron force and its force gradient as functions of the distance from the plate to the sphere. It can be found that the force approaches zero as the sphere get closer to the plate. This phenomena is in good agreement with the quadratic coupling of the symmetron to matter. A sensitive force gradient detection would be achieved if we trap the sphere in a relatively closed location, since the value of gradient would be higher.

\section{Numerical results}\label{3}
\subsection{The detection of the frequency shift}\label{3.1}

In this section, we aim to present practical parameters and numerical results for the predicted cavity detection. Initially, we set the pump-cavity detuning $\Delta_{p} = \omega_{cav} - \omega_{p} = 0$ for convenience. For the optical setup, we choose the trapping laser beam at the power P=0.17W with the wavelength $\lambda_\text{{laser}}=1064$nm injected into a cavity \cite{Delić_2020}. The pump beam is set to a power P=13nW at the frequency $\omega_p=2\pi\times4.77$GHz \cite{Jiang:20}.The cavity in our scheme is chosen to have an optical quality factor $Q\sim 10^{10}$ \cite{Gieseler_2013} with cavity amplitude decay rate $\kappa=2\pi\times215$kHz \cite{article3}. Two mirrors of the cavity are considered to be separated by a distance L=2mm, with the mirror served as source mass being flat while the radius of curvature for the other mirror is $R_\text{mirror}=10\mu$m in reference to settings in ref.~\cite{Hunger_2010}. With the above parameters, the mode volume  $V_c=1.09\times10^{-13}$m$^3$ can be calculated via $V_c\approx\pi L w_0^2/4$,  where $w_0$ is the waist for the fundamental mode. In addition, we can derive the coupling strength $g\sim 0.41$kHz with the following formula\cite{doi:10.1073/pnas.0912969107}
\begin{eqnarray}
g=\frac{3}{4}\frac{V_{sphere}}{V_c}\frac{\epsilon-1}{\epsilon+2}\omega_{cav},
\label{eq:ad1}   
\end{eqnarray}
where $\epsilon=3.75$ is the relative permittivity for fused silica and $V_{sphere}=4.2\times10^{-21}$m$^3$ . Then we can derive the Rabi frequency for pump laser as $\Omega_p\approx 2\pi$MHz considering a power density $\approx 10$mW/cm$^2$ \cite{9452539}. To achieve the mechanical damping rate, we choose a typical oscillator system with $\omega_{n} = 2\pi\times125$ kHz along the coupling direction \cite{Gieseler_2013}, leading to a gas-induced decay damping rate $\Gamma_{n} \approx 7.98 \times 10^{-7}$Hz decided by
\begin{eqnarray}
\Gamma_n=\frac{16p}{\pi \Bar{v}_{gas} R\rho_{sphere}},
\label{eq:ad2}   
\end{eqnarray}
where p=10$^{-10}$mbar is the gas pressure of vacuum cavity and mean velocity of gas is given by $\Bar{v}_{gas}=\sqrt{k_BT/m_{gas}}\approx 290m/s$ \cite{doi:10.1073/pnas.0912969107,Hunger_2010}. This allows the system to have a longer lifetime for potential operation, and this helps to ignore the effect of frequency drifts during measurement.

\begin{figure}[ht!]
	\centering
	\subfigure{
		\includegraphics[width=7cm]{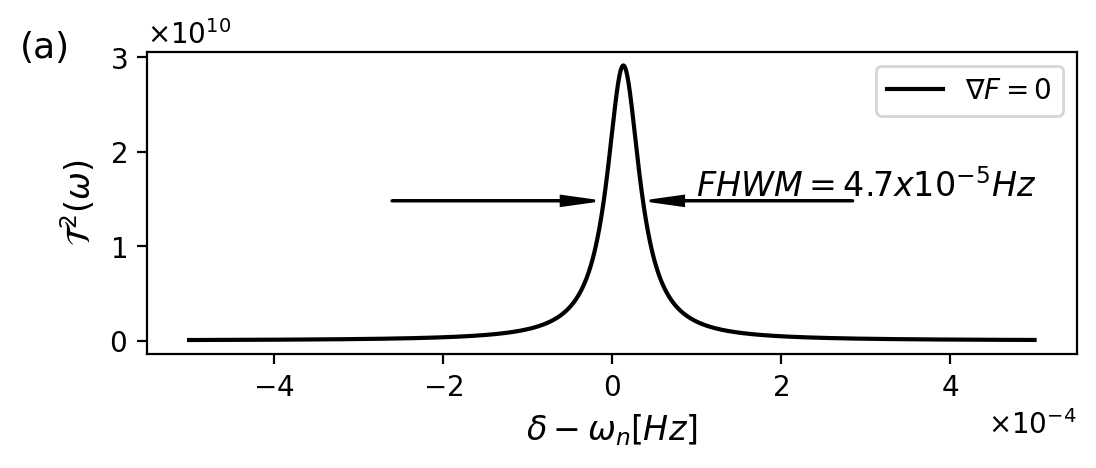}
        \label{fig:3-1}
        \vspace{0.02mm}
}
    \subfigure{
        \centering
		\includegraphics[width=7cm]{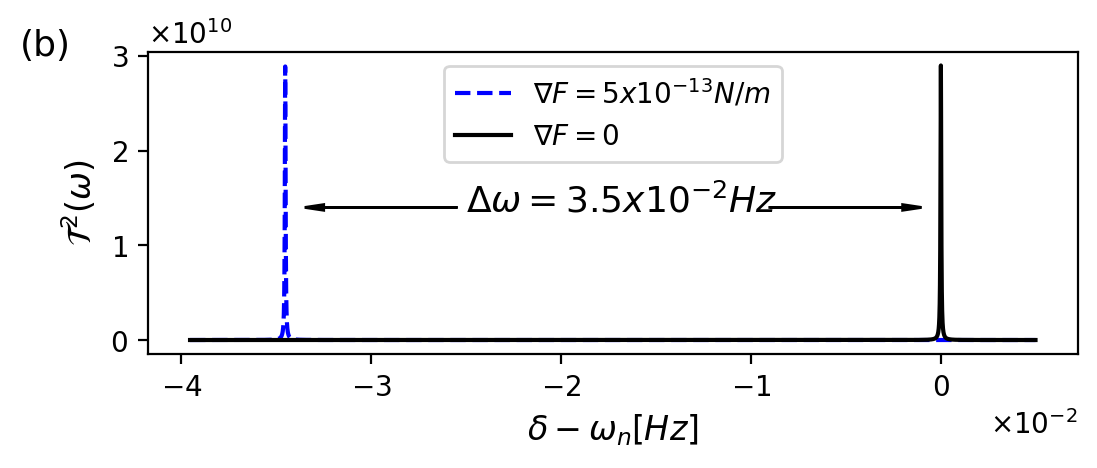}
        \label{fig:3-2}
}
\caption{\label{fig:3}Transmission $\mathcal{T}^2{(\omega)}$ of the probe field as a function of $\delta - \omega_n$. (a) A sharp peak can be found at $\delta-\omega_n$ without the presence of external force gradient. Additionally, the FWHM=4.7$\times10^{-5}$Hz is marked, representing the minimum distinguishable frequency discrepancy. (b) A frequency shift of $3.5\times10^{-2}$Hz is observed from the black line(before) to the blue line(after) with a gradient of $5\times10^{-13}$ N/m as an example. }
\end{figure}

To determine the ability of detection of our system, we should first investigate the probe spectrum through Eq.~(\ref{eq:24}). Shown in Fig.~\ref{fig:3-1}, the peak is located at the original frequency $\delta =\omega _n$ in the condition that there is no the fifth-force gradient($\nabla F=0$). The resolution of the system depends on the full width at half maximum (FWHM) of the oscillation peak, which in our scheme can be measured as 4.7$\times 10^{-5}$Hz. Fig.~\ref{fig:3-2} shows the impact of a nonzero force gradient on the spectrum. For instance, in the presence of a force gradient$\nabla F=5\times 10^{-13}$N/m, a frequency shift of $\Delta \omega =3.5\times 10^{-2}$Hz can be observed, which indicates that the effect of force gradient leads to a decrease of the resonance frequency. The relation between the frequency shift and force gradient is given by Eq.~(\ref{eq:nine}), by which we can derive the detection limit of our system is 

\begin{eqnarray}
\nabla F_{\text{min}}=\Delta\omega_{\text{FWHM}}\frac{2k}{\omega_n}= 6.80\times 10^{-16}N/m,
\label{eq:ad3}   
\end{eqnarray}
with $m_{eff}=9.2\times 10^{-18}$kg and $k=m_{eff}\omega _n^2=5.68\times 10^{-6}$N/m.

To get the precise minimum measurable frequency shift, we have to consider the physical noise during the detection process. Firstly, the fundamental limit of thermomechanical noise should be considered, originating from the thermally driven random motion of the realistic mechanical system. The noise is related to the measurement by its spectral density $S(\omega )$ and measurable bandwidth $\Delta f$\cite{10.1143/PTP.49.1516}. For a one-dimensional harmonic oscillator in our system, the spectral density of displacement $S_x(\omega)$ is given by\cite{10.1063/1.1642738}
\begin{eqnarray}
S_x(\omega) = \frac{S_F(\omega )}{m_{eff}^2 ((\omega^2 - \omega_n^2)^2 + \omega^2 \omega_n^2/Q^2)}.
\label{eq:0.1}   
\end{eqnarray}

\begin{figure}[ht!]
	\centering
	\subfigure{
		\includegraphics[width=7cm]{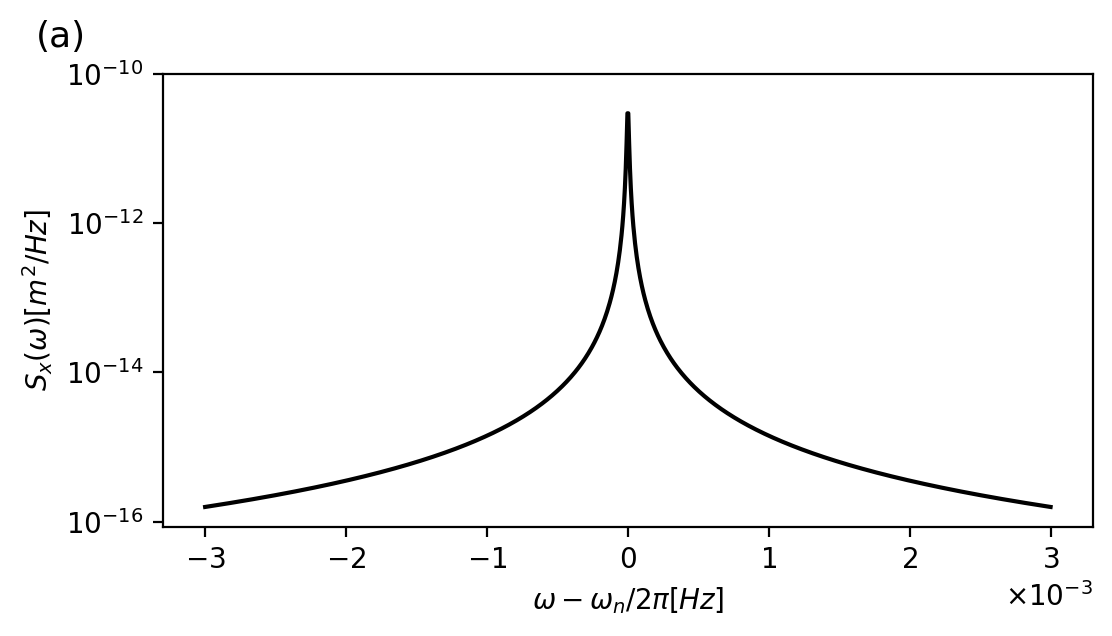}
        \label{fig:4-1}
        \vspace{0.02mm}
}
    \subfigure{
		\centering
		\includegraphics[width=7cm]{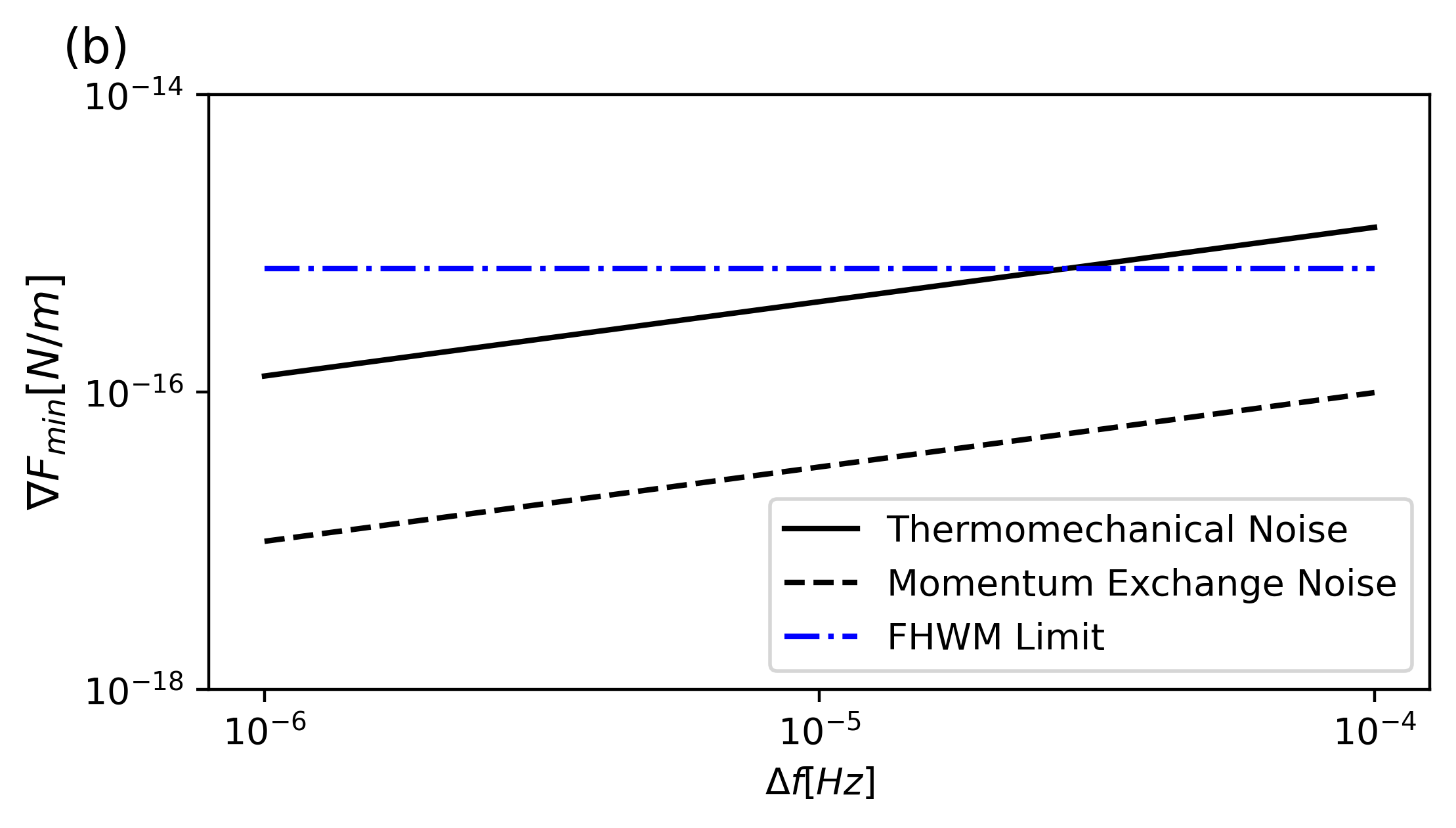}
        \label{fig:4-2}
}
\caption{\label{fig:4}
(a) The power spectral density of thermomechanical fluctuations as the function of the detuning to resonator frequency $\omega_n$. (b) Sensitivity limits due to  thermomechanical fluctuations and momentum exchange noise compared to the detection limit from FWHM. The elimination of the momentum exchange noise is considered under the condition $p=10^{-10}$mbar. With the main noise limit staying below the detection result, we can ignore their influence on the system safely.}
\end{figure}
The force spectral density here is given as $S_F(\omega) = 4 m_{eff} \omega_n k_B T / Q$ with $k_B=1.38\times10^{-23}$J/K as the Boltzmann’s constant and $T=293$K as the room temperature. In Fig.\ref{fig:4-1}, we illustrate the power spectral density of thermomechanical fluctuations to the  frequency. Since $\Delta f$ is in relation to the characteristic response time $\tau =Q/2\omega $ by $\Delta f \approx 1/2\pi \tau $, we can make an estimate that the maximum measurable bandwidth $\Delta f \approx 10^{-5}$Hz. The limited minimum frequency shift can be calculated by integrating the spectral density of the frequency fluctuations $S_{\omega}(\omega) = (\omega_n/2Q)^2 \cdot (S_x(\omega)/\langle x^2_{\text{rms}}\rangle)$ through the area in $\omega_n \pm \pi \Delta f$\cite{10.1063/1.1499745,Robins1984PhaseNI}. Finally, we get
\begin{eqnarray}
\Delta\omega_n \approx \sqrt{\frac{k_B T \Delta f }{m_{eff}\omega_n \langle x_{\text{rms}}\rangle^2 Q}},
\label{eq:0.2}   
\end{eqnarray}
here $\langle x_{\text{rms}} \rangle = \sqrt{k_B T/m_{eff} \omega_n^2}$ represents the root-mean-square (rms) amplitude of a trapped nanosphere at thermal equilibrium.

Besides the thermomechanical noise, the momentum exchange noise from the interaction between gas molecules and resonator is another attractive noise source. According to the discussion in ref.~\cite{10.1063/1.1642738}, we derive the spectral density $S_x(\omega)$ here with a similar form to the case of thermomechanical noise

\begin{eqnarray}
S_x(\omega) = \frac{S_F(\omega )}{m_{eff}^2 ((\omega^2 - \omega_n^2)^2 + \omega^2 \omega_n^2/Q_{gas}^2)}.
\label{eq:0.3}   
\end{eqnarray}
The quality factor considering gas dissipation is defined as  $Q_{gas} = m_{eff} \omega_n v / pA$, with $v = \sqrt{k_B T / m}$ is the thermal velocity of the gas molecules, $p$ the gas pressure, and $A = 4\pi R^2$ the surface area of the nanosphere. $\Delta \omega$ can be calculated in the same way to Eq.~(\ref{eq:0.2}). And it can be found in the equation that $\Delta \omega$ is proportional to the pressure $p$, indicating we can obtain higher sensitivity at higher vacuum level.

According to Eqs.~(\ref{eq:nine}) and (\ref{eq:0.2}), we have the expression of $\nabla F$ in this section as follows
\begin{eqnarray}
\nabla F= \sqrt{\frac{4k_B T \omega_n m_{eff} \Delta f}{\langle x_{rms} \rangle ^2Q}}.
\label{eq:0.4}   
\end{eqnarray}
To be specific, we plot $\nabla F$ as a function $\Delta f$ in the Fig.~\ref{fig:4-2}, showing the limit of different source of noise. Note that the limit of the dominant thermomechanical noise$\nabla F \sim 10^{-16}$N/m is lower than that of the FWHM calculation, allowing us to ignore the impact of this part of noise on the system. As for the exchange momentum noise, it is expected to be much lower in high-vacuum operation, resulting in an insignificant source of noise.

\subsection{Constraints on symmetron fields}\label{3.2}
Ideally, one can place constraints on the parameters of the symmetron force with the results in Sec.~\ref{3.1} and Eq.~(\ref{eq:1.6}). Before further discussing the symmetron force constraints, it is essential to consider the window of feasible $\mu$ value given by the dimension of the sphere-plate system. The upper bound of $\mu$ is related to the length of the cavity L along the resonant direction, considering the fact that only when $\mu \geq L^{-1} $ can the symmetron field have non-vanishing value in the cavity\cite{PhysRevD.91.063503,PhysRevLett.110.031301}. For the lower bound of $\mu$, it should meet the condition that $\mu R \ll 1$\cite{PhysRevD.101.064065}, otherwise the reaction of the sphere on the plate-induced symmetron field will be non-negligible. Recall the system we use in Fig.~\ref{fig:2}, the radius of the sphere is $R_{sphere}$=100nm and the dimension of vacuum chamber is L=2mm. Consequently, we can only set constraints of other parameters of the symmetron force within the regime $10^{-1}$eV $\sim 10^{-4}$eV.

With the definite range values of $\mu$, we can establish the constraints on the self-coupling constant M and $\lambda$ for the next step. Firstly, we turn to the relation of parameter M and $\mu$, considering the limit of background matter density. To ensure the calculation in Sec.~\ref{2.2} remain valid, the critical density $\rho_{crit}=\mu^2 M^2$ should satisfy: 
\begin{eqnarray}
\rho_{vac} < \mu^2 M^2 < \rho_{sphere}.
\label{eq:f1}   
\end{eqnarray}
In the condition of this paper, the bounds are $\rho_{vac}\approx 10$eV and $\rho_{sphere}\approx  10^{18}$eV in natural units, corresponding to the density of air under $10^{-10}$mbar at room temperature and the density of fused silica.

Additionally, to meet the condition that the sphere and the plate are strongly screened, $\lambda_{sphere} \ll 1$ is required. According to the form $\lambda_{sphere}=3M^2/\rho R^2$ from Eq.~(\ref{eq:1.3}), the constraint is written as 
\begin{eqnarray}
\frac{3M^2}{\rho_{sphere}R^2} < 1.
\label{eq:f2}   
\end{eqnarray}

\begin{figure}[ht!]
\centering
\includegraphics[width=10cm]{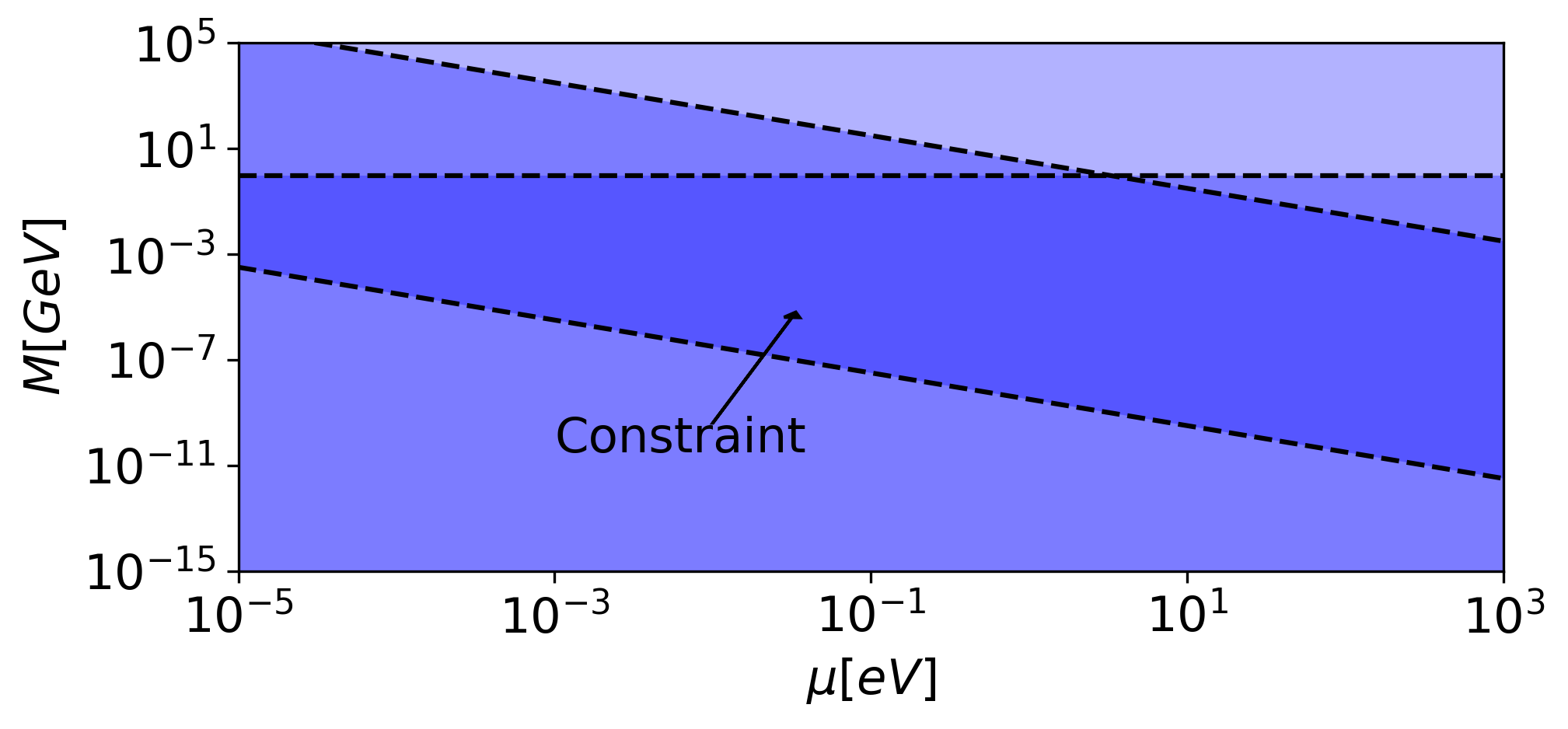}
\caption{\label{fig:5} Constraints of $M$ and $\mu$ derived from Eqs.~(\ref{eq:f1}) and (\ref{eq:f2}).The horizontal line refers to the requirement that the sphere is strongly screened from Eq.~(\ref{eq:1.3}).The slanted lines align with the condition that the symmetron field have non-zero value only in the vacuum cavity considering Eq.~(\ref{eq:two}). Otherwise the spontaneous symmetry breaking will occur inside the matter, or the field will stay zero for whole system, rendering the calculation to be irrelevant.}
\end{figure}
With Eqs.~(\ref{eq:f1}) and (\ref{eq:f2}), we illustrate the constraints in Fig.~\ref{fig:5}. The horizontal line corresponds to the result in Eq.~(\ref{eq:f2}), while the other two lines refer to the limits on M imposed through Eq.~(\ref{eq:f1}).And the center area filled with dark blue is the constraint we obtain.

The final step toward a complete constraint of symmetron force is to determine the range of $\lambda$. The forecast constraint on $\lambda$ is connected to our optomechanical detection by Eq.~(\ref{eq:1.6}). Notice that a higher force gradient $\nabla F$ indicate 
a smaller value of $\lambda$ for fixed $\mu$ and M. Thus, one can rule out the values of $\lambda$ which lead to a higher $\nabla F$ than the limit obtained in Eq.~(\ref{eq:ad3}). Besides, $\lambda > 0$ is necessary to make the symmetron mechanism to work. 

\begin{figure}[ht!]
	\centering
	\subfigure{
        \centering
        \includegraphics[width=0.45\textwidth]{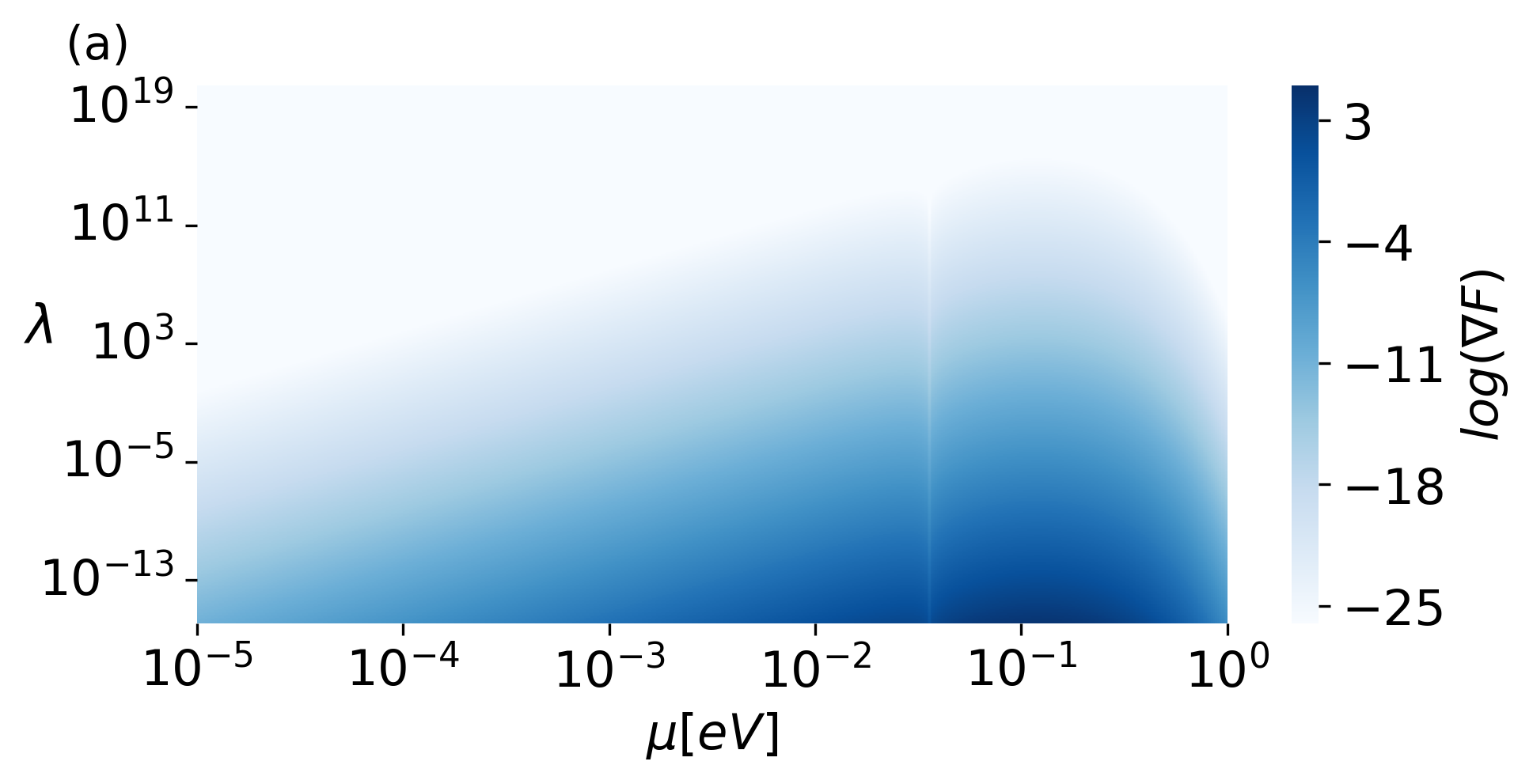}
        \label{fig:6-1}
}
    \subfigure{
		\centering
		\includegraphics[width=0.45\textwidth]{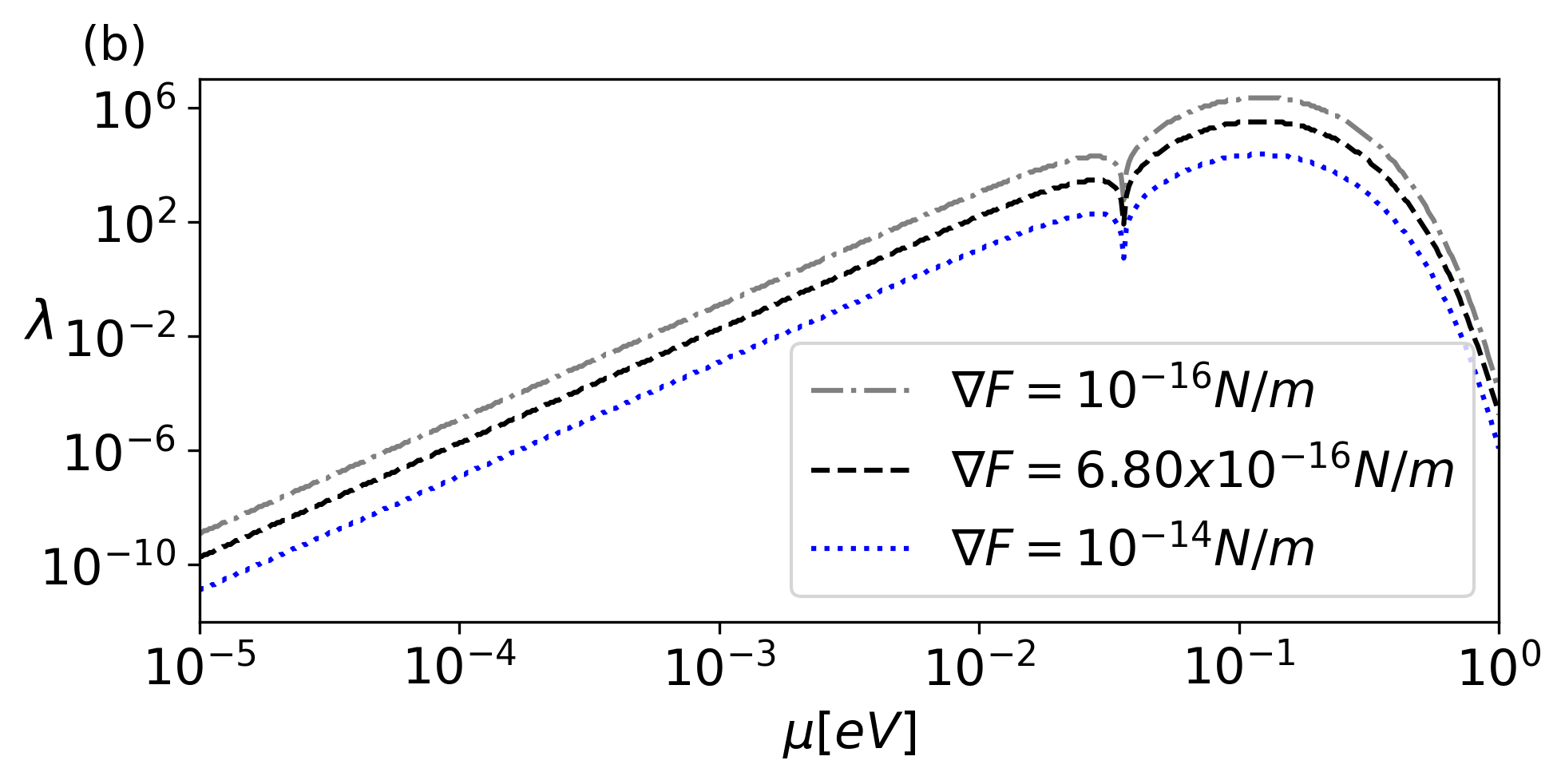}
        \label{fig:6-2}
}
\caption{\label{fig:6}(a) Symmetron-induced force gradient as the function of $\mu$ and $\lambda$, with the detect sphere of R=100nm located at d=5$\mu$m from the surface of source mass. (b) Forecast constraints on symmetron parameters $\mu$ and $\lambda$ according to the minimum $\nabla F$ in Sec.~\ref{3.1}. The three curves correspond to three bounds given by different sensitivities of $\nabla F=10^{-16}N/m,6.80\times10^{-16}N/m,10^{-14}N/m$ from the top to the bottom. Recall that the value of $\lambda$ which indicates larger $\nabla F$ than the result in Eq.~(\ref{eq:ad3}) will be ruled out. Namely, the area under the bound given by the black dashed line will be excluded.}
\end{figure}

In Fig.~\ref{fig:6-1}, we illustrate the value of force gradient $\nabla F$ with different values of M and $\mu$. Considering the data points obtained, a curve indicating the result achieved with optomechanical detection is plotted in  Fig.~\ref{fig:6-2}. The value of $\lambda$ under the curve will be ruled out. It should be emphasized that the parameter discussed are constrainable only in range $10^{-1}$eV $\sim 10^{-4}$eV. Consequently, we can finally establish a comprehensive area of parameters to exclude with the discussion in Sec.~\ref{3.2}. 

\begin{figure}[ht!]
\centering
\includegraphics[width=0.9\textwidth]{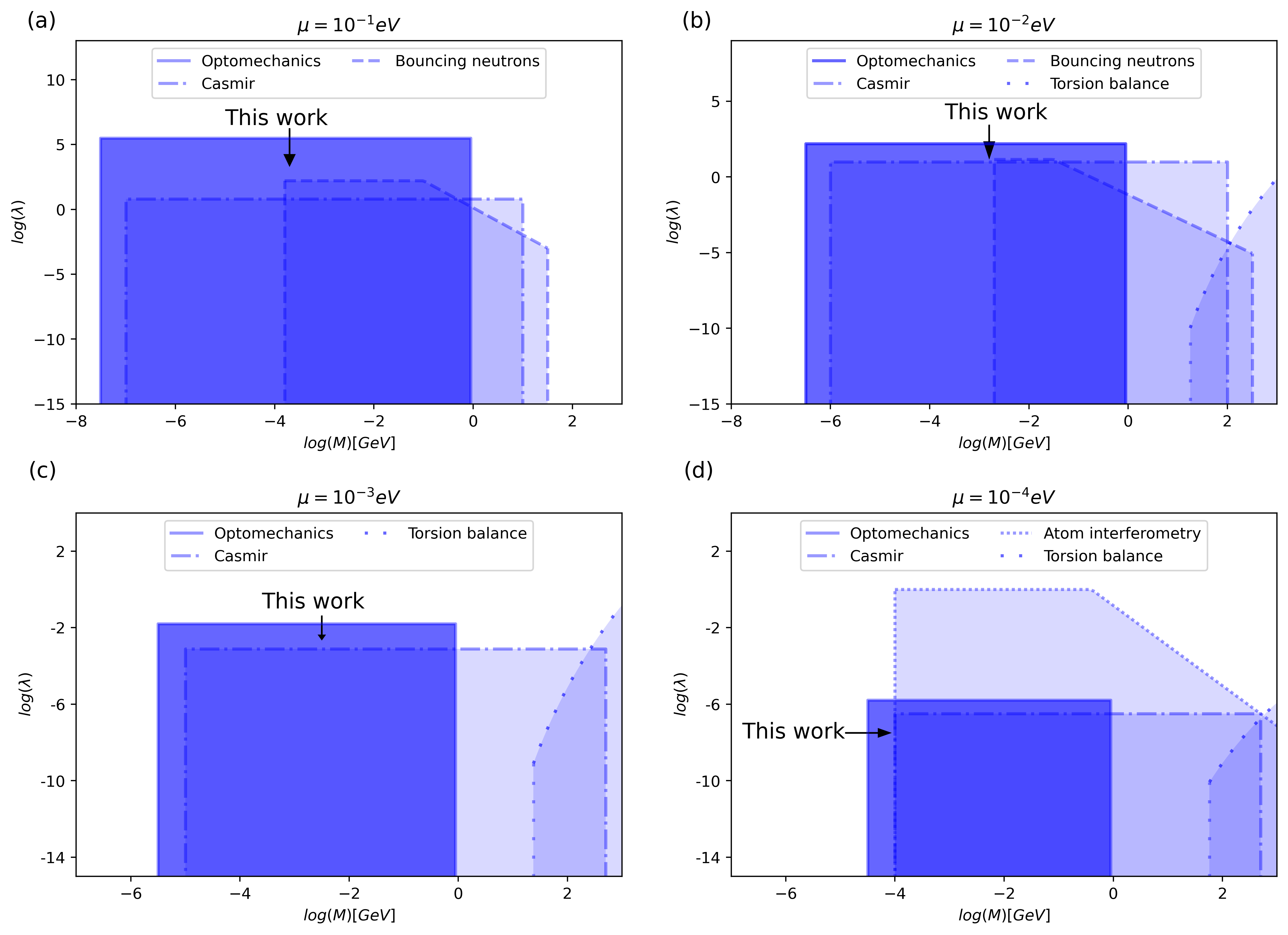}
\caption{\label{fig:7}Symmetron constraints from our hypothetical experiment scheme with other bounds in Casimir-force design, atom interferometry, torsion balance, and ultracold bouncing neutron measurements reproduced from refs. \cite{PhysRevD.101.064065,PhysRevLett.123.061102,PhysRevLett.110.031301,article} respectively. The excluded area of our method is based on the result in Figs.~\ref{fig:5} and \ref{fig:6} in range of $10^{-1}$eV $\sim 10^{-4}$eV, and shown within four discrete values $\mu=10^{-1}eV, 10^{-2}eV, 10^{-3}eV, 10^{-4}eV$.}
\end{figure}

In Fig.~\ref{fig:7}, we plot the constraints with discrete value of $\mu$. Our work is highlighted in dark blue while other constraints are filled with light blue, including an unperformed Casimir-force design\cite{PhysRevD.101.064065}, atom interferometry\cite{PhysRevLett.123.061102}, torsion balance\cite{PhysRevLett.110.031301}, and ultracold bouncing neutron experiments\cite{article}. Our hypothetical scheme is particularly effective for high value of $\mu$, which is found to exclude larger area compared to other methods. One can see that torsion balance method provides extra bounds in $M >10$Gev regime and atom interferometry gives strong constraints for $\mu=10^{-4}$eV, which are largely complementary to the force-based detection results such as this work and the Casimir-force experiment. Additionally, the area $M \sim 1$Gev is well constrained by the bouncing neutrons experiment, which has the potential to expand its constraint up to $\mu \sim $eV. Similar to our method, the Casimir-force constraints are provided through the minimum different force $\delta F$ of the symmetron field in a vacuum chamber at two locations\cite{PhysRevLett.116.221102}. This method shows a good ability to cover an extended range of $\sim$ 7 orders of magnitude. Taking all the constraints together into account, we can find that a substantial area in parameter space has been excluded. 

\section{Conclusion}\label{4}
In this paper, an optomechanical method is developed to set constraints on the symmetron force. In a system consisting a dense but small sphere and a relatively large plate, we demonstrate the behavior of the symmetron field with the screening mechanism. Via the theory of quantum optics, we derive the solution of the transmission spectrum and the function of associated noise in detection process. Further, the relation between symmetron-force-induced frequency shift and the optical measurable parameters is established, which leads to the minimum detectable force gradient. Based on results of the hypothetical experiment in our scheme, we deduce constraints on the symmetron force parameters, which make significant progress particularly for $\mu=0.1$eV of 3 orders of magnitude. Besides, there is still some work to do for future experiments and numerical simulation. With high-finesse cavities with larger volume and the method of cavity-assisted cooling \cite{PhysRevA.91.013824,PhysRevLett.118.183601,PhysRevD.98.030001,PhysRevA.64.033804}, hyperfine spectrum and more sensitive detection can be achieved. More precise numerical relaxation techniques can be introduced in force calculations, which can be found in refs\cite{article2,PhysRevD.94.044051}. And approximations for complex geometries could be helpful for more reliable constraints\cite{PhysRevD.76.124034,PhysRevLett.98.050403}. With the progress of experimental technology, we believe that there will be more actual optomechanics-based detection devices in the near future.

\acknowledgments
This work is supported by Natural Science Foundation of Shanghai (Grant No. 20ZR1429900).

% Bibliography

%% [A] Recommended: using JHEP.bst file
\bibliographystyle{JHEP}
\bibliography{biblio.bib}

\end{document}